\documentclass[letterpaper,superscriptaddress,prl,twocolumn,floatfix]{revtex4}

\RequirePackage{graphicx}
\newcommand{\D}{\mathrm d}
\newcommand{\I}{\mathrm i}

\begin{document}

\title{Coexistence of Weak and Strong Wave Turbulence in a Swell Propagation.}

\date{\today}

\author{V.\,E.~Zakharov}
\email{zakharov@math.arizona.edu}
\affiliation{Department of Mathematics, University of Arizona, 617 N. Santa Rita Ave., P.O. Box 210089, Tucson, AZ 85721-0089, USA}
\affiliation{P.\,N.~Lebedev Physical Institute RAS, 53 Leninsky Prosp., GSP-1 Moscow, 119991, Russian Federation}
\affiliation{Waves and Solitons LLC, 918 W. Windsong Dr., Phoenix, AZ 85045, USA}
\affiliation{L.\,D.~Landau Institute for Theoretical Physics RAS, 2 Kosygin Str., Moscow, 119334, Russian Federation}

\author{A.\,O.~Korotkevich}
\email{kao@itp.ac.ru}
\affiliation{L.\,D.~Landau Institute for Theoretical Physics RAS, 2 Kosygin Str., Moscow, 119334, Russian Federation}

\author{A.~Pushkarev}
\email{andrei@cox.net}
\affiliation{P.\,N.~Lebedev Physical Institute RAS, 53 Leninsky Prosp., GSP-1 Moscow, 119991, Russian Federation}
\affiliation{Waves and Solitons LLC, 918 W. Windsong Dr., Phoenix, AZ 85045, USA}

\author{D.~Resio}
\affiliation{Coastal and Hydraulics Laboratory, U.S. Army Engineer Research and Development Center, Halls Ferry Rd., Vicksburg, MS 39180, USA}


\begin{abstract}
By performing two parallel numerical experiments --- solving the
dynamical Hamiltonian equations and solving the Hasselmann kinetic
equation --- we examined the applicability of the theory of weak
turbulence to the description of the time evolution of an ensemble
of free surface waves (a swell) on deep water. We observed
qualitative coincidence of the results.

To achieve quantitative coincidence, we augmented the kinetic
equation by an empirical dissipation term modelling the strongly
nonlinear process of white-capping. Fitting the two experiments,
we determined the dissipation function due to wave breaking and
found that it depends very sharply on the parameter of
nonlinearity (the surface steepness). The onset of white-capping
can be compared to a second-order phase transition. This result
corroborates with experimental observations by Banner, Babanin,
Young \cite{BBY2000}.
\end{abstract}

\maketitle

\section{Introduction.}
Wave turbulence is realized in plasmas, liquid helium,
magnetohydrodynamics, nonlinear optics, etc. A perfect example of
wave turbulence is a wind-driven sea. The major conceptual
difference between wave turbulence and ``classical'' turbulence in
an incompressible fluid is the presence of a characteristic
dimensionless parameter $\mu$, characterizing the level of
nonlinearity. Turbulence is considered to be ``weak'' if $\mu \ll
1$, otherwise it is ``strong''. In classical hydrodynamic
turbulence $\mu \rightarrow \infty$.

A more rigorous definition of weak turbulence is the following:
this is the turbulence which is well described by the kinetic
equation for waves. These equations are the quantum kinetic
equations for bosons in the limit of very high occupation numbers.
They were derived in statistical physics in the late twenties
\cite{Nordheim1928, Peierls1929} and rediscovered in nonlinear
wave dynamics in the sixties. The kinetic equation, describing
four-wave resonant interaction of gravity waves, was named after
K.~Hasselmann, who derived it in 1962-1963 \cite{Hasselmann1962}.

The theory of weak turbulence is well-developed \cite{ZFL1992}.
The kinetic equation has rich families of Kolmogorov-Zakharov (KZ)
and self-similar solutions, which can be efficiently used for
explaining a wide range of experimental data
\cite{Zakharov2005,BPRZ2005}. However, today we have a clear
understanding of the following fact: even for small values of
$\mu$, the theory of weak turbulence may be incomplete. In many
important physical situations weak and strong turbulence coexist.

Even if the weak turbulent resonant interaction effects dominate
in the greater part of space, strongly nonlinear effects could
appear as rare localized coherent events. If they are smooth and
regular, they are solitons, quasisolitons or vortices. However,
they could be catastrophic, in which case they are wave collapses,
similar to self-focusing in nonlinear optics or Lagmuir collapses
in plasma. Even rare sporadic collapse events can essentially
affect the physical picture of wave turbulence.

There are two main types of wave collapse events in a wind-driven
sea. The first is the formation of freak waves; this is not a
subject of our study. The second, which is much more common, is
wave-breaking or white-capping, which is an essential mechanism of
wave energy dissipation. It would be hopeless to develop an
efficient operational model of wave forecasting without an
understanding and a proper parametrization of this fundamental
effect. Meanwhile, a reliable analytical theory of this phenomenon
is still not developed, while field and laboratory experimental
data are scarce. The most promising approach to resolving this
problem is a massive numerical experiment.

The most informative experiment would be one that could provide a
direct numerical solution of the primitive dynamic equations
describing the wave ensemble. In 1992, Dyachenko, Pushkarev,
Newell and Zakharov numerically solved 2-D focusing NLSE and
observed the coexistence of self-focusing collapses with weak
turbulence \cite{DNPZ1992}. Later on, the 1-D MMT (Maida,
McLaughlin and Tabak) model and its generalizations were solved
numerically by different authors (see summary in \cite{DPZ2004}).
Again, the coexistence of wave collapses and weak turbulence was
verified. In our article, we present (as we hope, for the first
time in the literature) the results of a far more detailed
experiment. We performed the numerical simulation of the evolution
of an ocean swell using two different approaches.

In the first, we solved the Euler equations for the 3-D potential
flow of an ideal incompressible fluid with a free surface in the
presence of gravity. We used the Hamiltonian form of these
equations \cite{Zakharov1968, Zakharov1999}. For gravity waves,
the parameter of nonlinearity is the average steepness $\mu$. We
expanded the Hamiltonian in powers of $\mu$ up to order $\mu^4$.
In the second experiment, we solved the Hasselman kinetic
equation.

The comparison of the results demonstrates qualitative accordance.
Both experiments describe expected effects, such as the downshift
of the spectrum peak, the angular spreading of the spectrum and
the formation of Zakharov-Filonenko spectral tails $F_\omega \sim
\omega^{-4}$ \cite{Zakharov1966, ZFL1992}. To obtain quantitative
coincidence of the results, we have to augment the Hasselmann
equation by an empirical dissipation term $S_{diss}$, modelling
white-capping effects. We tried several versions of this term. The
versions of $S_{diss}$ used in the industrial wave-predicting
models WAM3 and WAM4 essentially overestimate the dissipation for
a moderate steepness. The comparison with dynamical computations
shows that white-capping dissipation decreases dramatically with
decreasing steepness and that it is probably a threshold
phenomenon, similar to a second-order phase transition. Similar
results were earlier obtained in the field experiment by Banner,
Babanin and Young \cite{BBY2000}.

\section{Dynamical model}
In this part of our experiment, the surface of the liquid is
described by two functions of the horizontal variables $x,y$ and
the time $t$: the surface elevation $\eta(x,y,t)$ and the velocity
potential on the surface $\psi(x,y,t)$. In our approximation, they
satisfy the following equations \cite{Zakharov1968}:
\begin{equation}
\label{eta_psi_equations}
\begin{array}{rl}
\displaystyle
\dot \eta = &\hat k  \psi - (\nabla (\eta \nabla \psi)) - \hat k  [\eta \hat k
\psi] +\\
\displaystyle
&+ \hat k (\eta \hat k  [\eta \hat k  \psi]) + \frac{1}{2} \nabla^2 [\eta^2 \hat
k \psi] +\\
\displaystyle
&\frac{1}{2} \hat k [\eta^2 \nabla^2\psi] + \hat F^{-1}[\gamma_k \eta_k],\\
\displaystyle
\dot \psi = &- g\eta - \frac{1}{2}\left[ (\nabla \psi)^2 - (\hat k \psi)^2
\right] - \\
\displaystyle
&- [\hat k  \psi] \hat k  [\eta \hat k  \psi] - [\eta \hat k  \psi]\nabla^2\psi
+ \hat F^{-1}[\gamma_k \psi_k].
\end{array}
\end{equation}
Here $\hat k$ is the linear integral operator $\hat k
=\sqrt{-\nabla^2}$, $\hat F^{-1}$ corresponds to the inverse
Fourier transform.

Equations (\ref{eta_psi_equations}) are nowadays widely used in
numerical experiments and are solved by different versions of the
spectral code
\cite{PZ1996,Pushkarev1999,PZ2000,Tanaka2001,Onorato2002,DKZ2003grav,
DKZ2004, Yokoyama2004,ZKPD2005,LNP2006,Nazarenko2006,DKZ2003cap}.
In the present experiment, we solved the equations in the real
space domain $2\pi\times 2\pi$ using the finest currently possible
rectangular grid $512\times 4096$, putting $g=1$. The dissipative
terms $\hat F^{-1}[\gamma_k \eta_k]$ and $\hat F^{-1}[\gamma_k
\psi_k]$ are taken in the form of pseudo-viscous high frequency
damping. We put
\begin{equation}
\label{Pseudo_Viscous_Damping}
\begin{array}{l}
\displaystyle
\gamma_k = \left\{
\begin{array}{l}
\displaystyle
0, k < k_d,\\
\displaystyle
- \gamma (k - k_d)^2, k \ge k_d,\\
\end{array}
\right.\\
\displaystyle
k_d = 1024, \gamma = 5.65 \times 10^{-3}.
\end{array}
\end{equation}
In accordance with recent results \cite{DyachenkoDiasZakharov},
the dissipation term should be included in both equations.

The distribution of the wave action is described by the function
$n(k,t)=|a_{\vec{k}}(t)|^2$, where
\begin{equation}
a_{\vec k} = \sqrt \frac{\omega_k}{2k} \eta_{\vec k} + \I \sqrt
\frac{k}{2\omega_k} \psi_{\vec k},
\end{equation}
are complex normal variables. Here $\omega_k = \sqrt{gk}$.

As the initial condition, we used a Gaussian-shaped distribution
in the Fourier space:
\begin{equation}
\label{Dynamic_initial_conditions}
\begin{array}{l}
\displaystyle
\left\{
\begin{array}{l}
\displaystyle
|a_{\vec k}| = A_i \exp \left(- \frac{1}{2}\frac{\left|\vec k - \vec
k_0\right|^2}{D_i^2}\right),
\left|\vec k - \vec k_0\right| \le 2D_i,\\
\displaystyle
|a_{\vec k}| = 10^{-12}, \left|\vec k - \vec k_0\right| > 2D_i,
\end{array}
\right.\\
\displaystyle
A_i = 0.92\times10^{-6}, D_i = 60,\\
\displaystyle
\vec k_0 = (0; 300), \omega_0 = \sqrt{g k_0}.
\end{array}
\end{equation}
The initial phases of all harmonics were random. The average
steepness of this initial condition, defined as $\mu = \sqrt{2
\langle |\nabla\eta|^2\rangle}$, was $\mu \simeq 0.176$.

\begin{figure}[htb]
%\centering
\includegraphics[width=3.0in]{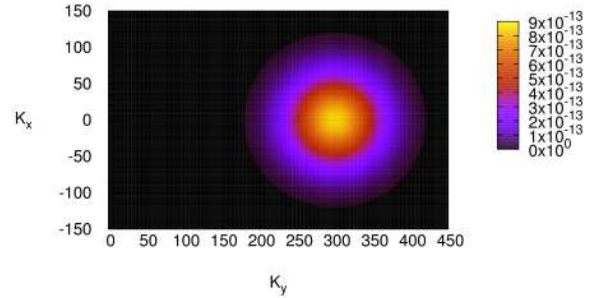}
\caption{\label{Spectrum_initial}Initial spectrum $|a_{\vec k}|^2$. $t=0$.}
\end{figure}

\begin{figure}[htb]
\centering
\includegraphics[width=3.0in]{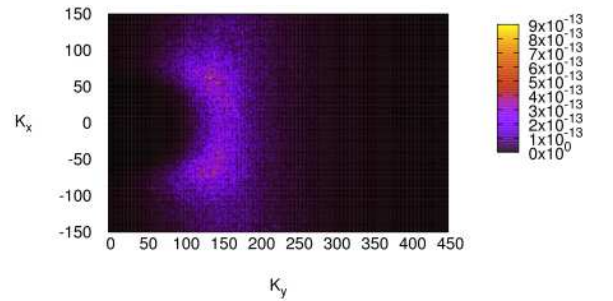}
\caption{\label{Spectrum_final}Final spectrum $|a_{\vec k}|^2$. $t \simeq 2809 T_0$.}
\end{figure}

\begin{figure}[htb]
\centering
\includegraphics[width=3.0in]{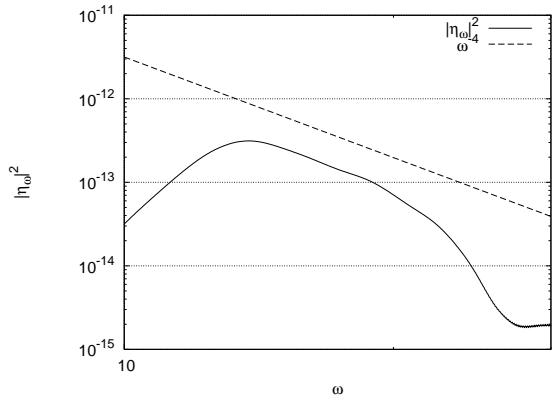}
\caption{\label{Kolmogorov_k}The time-averaged spectrum of a point
on the surface $n_{\omega} = \langle|\eta_{\omega}|^2\rangle$ in
the double logarithmic scale. The tail of the distribution fits to
the Zakharov-Filonenko spectrum $\omega^{-4}$.}
\end{figure}

\begin{figure}[htb]
\centering
\includegraphics[width=3.0in]{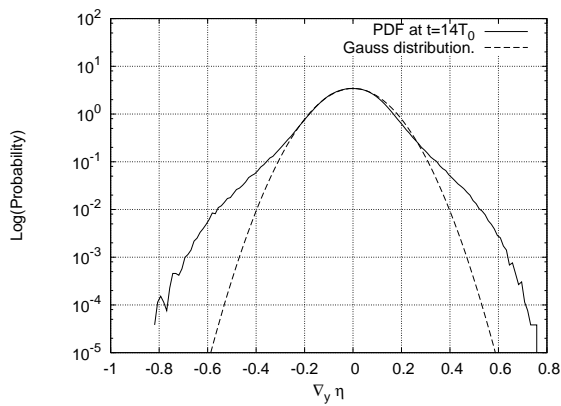}
\caption{\label{GradY_max}PDF for $(\nabla \eta)_y$ at the moment of the maximum
surface roughness. $t\simeq14T_0$.}
\end{figure}

The period of the most intensive wave was $T_0 = 2 \pi/\sqrt{300}
= 0.362$. Calculations continued until $t = 3378 T_0$. We observed
an angular spreading of the initial spectral distribution together
with a downshift of the spectral peak. Level-lines of the initial
and the final spectra are presented on
Figs.~\ref{Spectrum_initial}, \ref{Spectrum_final}.
We observed the following indications of wave-turbulent behavior:
\begin{enumerate}
\item{} The statistics of energy-capacity spectral modes is close
to the Rayleigh distribution. We observed the presence of a few
very intensive harmonics (so-called oligarchs, cf.
\cite{ZKPD2005}), which did not obey the Rayleigh statistics, but
their contribution to the total balance of the wave action is
small (no more than $5 \%$). This means that we almost overcame
negative effects caused by the finite size of our system (see
\cite{ZKPD2005,LNP2006,Nazarenko2006}), and that our grid is fine
enough. \item{} We observed the formation of the
Zakharov-Filonenko spectral tail in the energy spectrum
$|\eta_\omega|^2$ (see Fig.~\ref{Kolmogorov_k}).
\end{enumerate}
At the same time, we observed a manifestation of strong-turbulent
effects. They are manifested by the formation of ``fat tails'' on
the PDF for surface elevations and especially for its gradients
(see Fig.\ref{GradY_max})
The presence of these tails indicates that the surface has a
tendency to become rough and to produce white-capping. In our
model, wave-breaking is arrested by the strong pseudo-viscosity.

\section{Statistical experiments}
In the second experiment, we solved the Hasselmann kinetic
equation for $n_{\vec k} = \langle |a_{\vec k}|^2\rangle$
\cite{Hasselmann1962}
\begin{equation}
\label{Hasselmann_equation}
\begin{array}{l}
\displaystyle
\frac{\partial n_{\vec k}}{\partial t}=S_{nl}[n] + S_{diss} + 2\gamma_k n_{\vec k}, \\
\displaystyle
S_{nl}[n]=2\pi g^2 \int |T_{\vec k,\vec k_1,\vec k_2,\vec k_3}|^2 \left(n_{\vec k_1}n_{\vec k_2}n_{\vec k_3}+\right.\\
\displaystyle
\left. + n_{\vec k}n_{\vec k_2}n_{\vec k_3} - n_{\vec k}n_{\vec k_1}n_{\vec k_2} - n_{\vec k}n_{\vec k_1}n_{\vec k_3}\right)\times\\
\displaystyle
\times\delta\left( \omega_k+\omega_{k_1}-\omega_{k_2}-\omega_{k_3}\right)\times\\
\displaystyle
\times\delta\left(\vec k +\vec k_1 -\vec k_2 -\vec k_3\right)\,\D
\vec k_1\D\vec k_2 \D\vec k_3.
\end{array}
\end{equation}
Here $\gamma_k$ is the pseudo-viscosity and $S_{diss}$ is the
phenomenological dissipation term modelling the white-capping
process.

Eq. (\ref{Hasselmann_equation}) was solved on the grid $71 \times
36$ in polar coordinates on the frequency-angle plane by the
Resio-Tracy code \cite{RT1982}, improved in \cite{Zakharov2005,
BPRZ2005}. We first performed the experiment with $S_{diss}=0$. We
observed good qualitative coincidence with the dynamical
experiment. We observed a downshift of the spectral peak, angular
spreading and the formation of $\omega^{-4}$ spectral tails. But
the quantitative agreement of the experiments was not good: it was
clear that the inclusion of some phenomenological dissipation is
necessary.

We examined the standard from of $S_{diss}$ used in in the
industrial operational models of wave forecasting --- WAM Cycle 3
and WAM Cycle 4 (hereafter WAM3 and WAM4) \cite{SWAN}:
\begin{equation}
\label{WAMdissipation}
S_{diss}= C_{ds} \tilde{\omega} \frac{k}{\tilde{k}} \left((1-\delta)+\delta\frac{k}{\tilde{k}}\right)\left(\frac{\tilde{S}}{\tilde{S}_{pm}}\right)^p n_k
\end{equation}
where $k$ and $\omega$ are the wave number and the frequency,
tilde denotes the mean value; $C_{ds}$, $\delta$ and $p$ are
tunable coefficients; $S=\tilde{k}\sqrt{H}$ is the overall
steepness; $\tilde{S}_{PM}=(3.02\times 10^{-3})^{1/2}$ is the
value of $\tilde{S}$ for the Pierson-Moscowitz spectrum (note that
the characteristic steepness is $\mu \simeq \sqrt{2} S$). It is
worth noting that according to \cite{BBY2000}, the theoretical
value of the steepness for the Pierson-Moscovitz spectrum is
$S_{PM} \simeq (4.57\times 10^{-3})^{1.2}$, which gives us
$\mu\simeq 0.095$.

\begin{figure}[ht]
\centering
\includegraphics[width=3.0in]{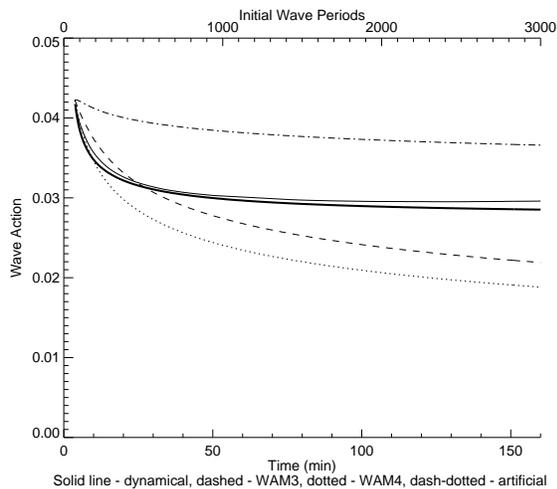}
\caption{Total wave action as a function of time. The solid line
corresponds to the dynamical equations, the dashed-dotted line ---
to the kinetic equation with artificial viscosity, the dashed line
--- to the kinetic equation with the WAM3 damping term, the dotted
line --- to the kinetic equation with the WAM4 damping term, and
the thick solid line
--- to the kinetic equation with the new damping term,}
\label{Action}
\end{figure}

The values of tunable coefficients in the WAM3 case are:
\begin{equation}
C_{ds} = 2.36 \times 10^{-5},\,\,\,\delta=0,\,\,\,p=4
\end{equation}
and in the $WAM4$ case are:
\begin{equation}
C_{ds} = 4.10 \times 10^{-5},\,\,\,\delta=0.5,\,\,\,p=4
\end{equation}
The evolution of the total wave action is presented on
Fig.~\ref{Action}. One can see that in the long run, the models
WAM3 and WAM4 overestimate white-capping dissipation. To achieve
better agreement of both experiments, we used the following form
of the dissipative term:
\begin{equation}
\label{proposed_coeff}
C_{ds} = 1.00 \times 10^{-6},\,\,\,\delta=0,\,\,\,p=12
\end{equation}
The total wave action curve corresponding to this new dissipation
term is shown on Fig.~\ref{Action} by the thick solid line and
displays excellent correspondence with the dynamical model.
\section{Conclusion}
Our experiments can be interpreted as a confirmation of the theory
of weak turbulence. However, even at moderate values of the
parameter of the nonlinearity $\mu$, the strongly nonlinear
effects of white-capping are essential. They manifest themselves
as fat tails of the $PDF$ and lead to additional dissipation of
wave energy. This dissipation demonstrates a very strong
dependence on the steepness. At steepness $\mu=0.176$ they
dominate, at steepness $\mu=0.09$ they are negligibly small. The
results of our experiments are in good qualitative agreement with
the field experiment of Banner, Babanin and Young \cite{BBY2000},
who found that wave-breaking is a threshold effect, similar to a
second-order phase transition.

We stress that the dependence (\ref{proposed_coeff}) is much
sharper than it is usually stated. So far, the sharpest dependence
$p=5$ was given by Donelan \cite{Donelan2001}. We can guess that
the real dependence of $S_{diss}$ on $\mu$ is even stronger, and
that the onset of the wave breaking is a threshold-type phenomenon
like a second-order phase transition.

\section{Acknowledgments}
This work was partially supported by ONR grant N00014-03-1-0648,
US Army Corps of Engineers grant DACW 42-03-C-0019 and by NSF
grant NDMS0072803, RFBR grant 06-01-00665-a, INTAS grant 00-292,
the Program ``Nonlinear dynamics and solitons'' from the RAS
Presidium and ``Leading Scientific Schools of Russia'' grant
NSh-7550.2006.2. A.O. Korotkevich was also supported by the
Russian President grant for young scientists MK-1055.2005.2.

The authors would also like to thank the creators of the
open-source fast Fourier transform library FFTW~\cite{FFTW} for
this fast, portable and completely free piece of software.

\end{document}